# A Debris Backwards Flow Simulation System for Malaysia Airlines Flight 370

Mike Eichhorn, Alexander Haertel
Institute for Automation and Systems Engineering
Technische Universitaet Ilmenau
98684 Ilmenau, Germany
Email: {mike.eichhorn, a.haertel}@tu-ilmenau.de

*Abstract*—This paper presents a system based on a Two-Way Particle-Tracking Model to analyze possible crash positions of flight MH370. The particle simulator includes a simple flow simulation of the debris based on a Lagrangian approach and a module to extract appropriated ocean current data from netCDF files. The influence of wind, waves, immersion depth and hydrodynamic behavior are not considered in the simulation.

## I. INTRODUCTION

The search for missing objects at sea or the detection of underwater pollution sources using all available information require robust algorithms. This research field is a highly topical subject after the crash of flight MH370. The discovery of its debris on the island of La Réunion at the end of July 2015 enables a search for the possible crash site using complex ocean simulations. Such ocean simulations are also necessary for the path planning of an AUV in time-varying ocean flows which is a research field of our group. For this planning, ocean prediction systems provide future ocean current data for a one to two week forecast window. These systems can also provide past ocean current data created by numerical models and adapted with real in situ measurements of drifting buoys, probes, measuring stations or satellite information. In [1] the flow fields from the HYCOM global ocean model were used to analyze possible debris positions of MH370 over two years.

The used approach to analyze possible crash sites in this work is based on a two-way Particle-Tracking Model (PTM). It is presented in detail in [2] and was carried out in a realistic hydrographic model over the East China Sea shelf for the period from June to August 2014.

This paper starts with a description of the two-way PTM and the concept of ocean current data extraction from netCDF files using interpolation methods. A short overview of the software framework is presented in the central part. The capability of the system will be demonstrated using an analytical ocean current model. Finally, the paper will discuss possible crash sites using four ocean models of HYCOM [3] and Copernicus Marine environment monitoring service (CMEMS) [4] in the two-way PTM.

## II. TWO-WAY PARTICLE TRACKING MODEL

### A. Flow Simulation Model

To simulate the possible location of the debris (particle) **x**=(*x,y*) in a current field **u**=(*u,v*), a Lagrangian approach for two-dimensional (2D):

$$\mathbf{x}^{t+\Delta t} = \mathbf{x}^{t+\Delta t} + \int_{t}^{t+\Delta t} \mathbf{u}(x,y,t)dt \quad (1)$$

was solved. A random walk process, which will typically be used in such an approach, was not considered. The inclusion of uncertain information occurs to the variability of the found site and time (see section V.C).

### B. ODE-Solvers

The time integral in equation (1) is solved using explicit one-step Runge-Kutta methods. We test several approaches from second order to fifth order to analyse their influence on the simulation accuracy. All approaches are embedded methods, which are designed to produce an estimation of the truncation error for the step size control.

The following table shows the chosen methods, the MATLAB function calls (if available) and additional information about the internal formulas (Butcher tableau, error calculation).

TABLE I. USED ODE SOLVERS

| Order *n* | Name | Information |
|---|---|---|
| 2 | Heun's method | |
| 3 | Bogacki-Shampine method (`ode23`) | [5] |
| 4 | Zonneveld method | [6] |
| 5 | Dormand–Prince method (`ode45`) | [6], [7] |

To determine the new step size *h* by using automatic step size control the following equation has to be solved [6]:

$$h = \max\left\{h_{min}, \min\left\{h_{max}, \tau h\left(\frac{\varepsilon_{tol}}{err}\right)^{\frac{1}{n}}\right\}\right\} \quad (2)$$

The parameter $\tau$ is a safety factor ($\tau \in (0, 1]$), usually $\tau$=0.8-0.9. The parameters $h_{min}$ and $h_{max}$ are the minimal and maximal step size. Acceptance or rejection of this step will depend on the error $err = \|\mathbf{x}_{err}\|$ to a defined tolerance $\varepsilon_{tol}$.

## C. Two-Way Particle-Tracking Model

To specify a possible crash site the two-way Lagrangian particle model (PTM) which is described in detail in [2] was used. Fig. 1 shows a schematic diagram to demonstrate the operating principle published in [2]. The first step (b) is the determination of source candidates (in this work, possible crash sites), generated by a backward-in-time PTM started from a receptor (in this work, the found site). In step two (c), particles are created from these sites to simulate a possible flow using a forward-in time PTM. The last step includes the analysis of the final particle positions for each source candidate. A rotated ellipse can describe the particle distribution, where the standard deviation $\sigma$ of the particles to the centre position corresponds with its minor- and major-axis. If the receptor is located within the $2\sigma$ ellipse, the source should be accepted as a possible true source (In this example it is source $S_0$).

In this work possible source candidates will be analysed using a histogram of the particle positions generated by a backward-in-time PTM starting from the found site. The selection of the true source takes place through the analyses of the average position of the particles and the corresponding histogram, because a normal distribution of the particle clusters after the simulation is not determined. Another difference to [2] is the omission of a random-walk process (see section V.C for details).

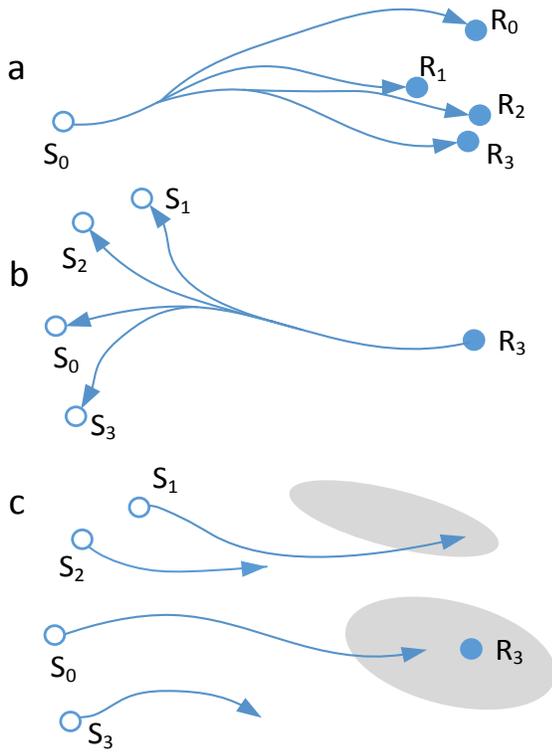

Fig. 1. Schematic diagram of two-way PTMs (a) Objects released from the true source $S_0$ reach four receptors $R_0$, $R_1$, $R_2$, $R_3$. (b) From the found site $R_3$ start a backward-in-time PTM to find source candidates $S_0$, $S_1$, $S_2$, $S_3$ (c) These several source candidates start a forward-in-time PTM. The gray marked ellipses show the distribution of the released particles from every source candidate. [2]

## III. CURRENT MODEL

### A. Interpolation

Since the ocean current data - coming from Ocean General Circulation Models (OGCM) as data files - will be provided only at discrete times and positions with a coarser time and length scale than required for particle track simulation, a multi-dimensional interpolation scheme will be utilized to extract the desired data. Fig. 2 outlines the scheme for the ocean current component $v$. For the interpolation Nearest-Neighbour, Linear, Cubic, Cubic Spline and Akima interpolation methods are available. The first interpolation step uses a two-dimensional interpolation function to extract the ocean current information for the several depth layers. The interpolation for depth and time are calculated separately using one-dimensional interpolation functions and the data produced by the depth-interpolation. Nearest neighbour and linear interpolation require two sampling points, meaning that two fields (for time $t$) or layers (for depth $z$) are required to determine the ocean current between two data points of an ocean model. All other methods use polynomials of order 3 to interpolate data and thus require at least four, ideally eight (Cubic Spline, Akima) or more fields or layers in order to generate the ocean current component $v$ at the defined position ($\mathbf{x_i}$, $y_i$) at the depth $z_i$ and at the time $t_i$.

A comparison between the available interpolation methods shows that the cubic spline interpolation recreates missing data of the used ocean models most precisely while also taking the most cpu time [8]. The implementation of the Akima interpolation [9] can make allowance for an abrupt change of ocean current conditions in case of tides or different depth streams. For time interpolation, the linear interpolation method was found to be a good compromise between precision and cpu time for data extraction.

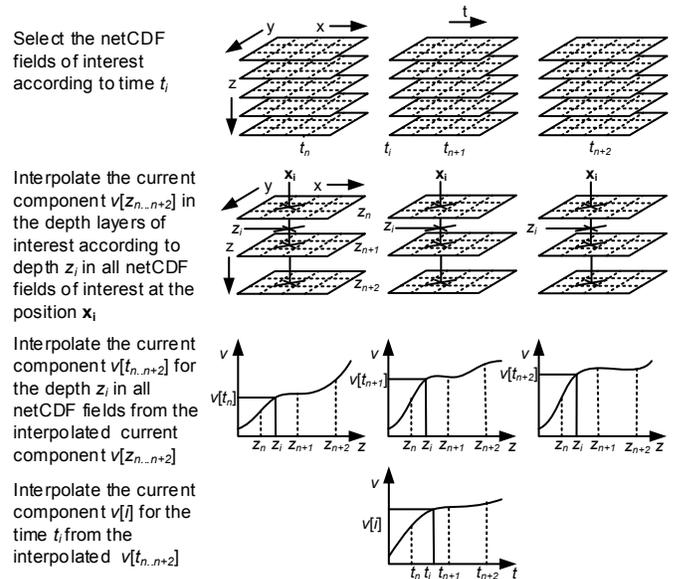

Fig. 2. Steps to interpolate an ocean current component $v[i]$ at position $\mathbf{x}[i]$, depth $z_i$ and time $t_i$ from the netCDF fields

## IV. Program Technical Implementation

This section provides an overview of some software technical details and the software products and libraries used. Fig. 3 outlines the various components of the flow simulation system with the software used.

The simulation system is written in C++ using the Microsoft Visual Studio 2013 [10]. This enables a fast development and an easy debugging. Make-files and shell-scripts exist in order to compile the code for Linux based system or Cygwin with the GNU Compiler Collection (GCC). The data, which includes the ocean current information of the eighteen-month drift, are available in netCDF file format provided by an OGCM. In this project, we use GLBa0.08 and GLBu0.08 of HYCOM [3] and the 2 hourly and 24 hourly global-analysis-forecast-phys-001-002 model of the Copernicus Marine Environment Monitoring Service (CMEMS) [4].

Since all these models have a different netCDF-structure a conversion to a custom netCDF-structure is necessary. A merging of ocean data is necessary because the prediction systems usually do not provide data of the time and area of interest in one piece. To do this, MATLAB or Python scripts can be used. If the data is stored in a different format, the FIMEX library [11] allows a conversion to netCDF. FIMEX is a File Interpolation, Manipulation, and Extraction library for gridded geospatial data, written in C/C++. It supports different data formats (currently netCDF, NcML, grib1/2 and felt), and allows their conversion from one format into another, to change the projection and interpolation of scalar and vector grids [11]. An XML file, which includes all necessary simulation parameters such as found site, found time, crash time, projection definitions (projection function, reference position, cartographic parameters) and settings are passed to the PTM and to the ODE solver. The Xerces-C++ XML Parser [12] is used to parse the files in several C++ programs.

The PTM generates a start position and a start time randomly for every particle (see section V.C) which is sent to the track simulator. With this information the main loop starts the ODE solver for a defined time range of 24 h. After this time the simulated end position will be returned to the main loop. This position will be logged and serves as new start point. This process will be repeated until the end time $t_{end}$ is reached. The original idea of this loop concept was the adaption (changing the tolerance $\varepsilon_{tol}$) of the ODE solver during the simulation to achieve a high stability and speedup. This will currently not be used. The available ODE solvers in the track simulator are presented in section II.B.

The interfaces of the ocean current model are based on a bridge design pattern [13]. In this pattern, there exist abstract C++ classes which define public functions. The derivations of these classes include the code of an implemented ocean current model. This enables an easy verification of the system using reference ocean current models such as the mathematical ocean model in section V.A. The standard function call of the ocean current model provides the two ocean current components $u$ and $v$ according to the defined position $\mathbf{x}$, depth $z$ and time $t$. It is also possible to receive additional information about their partial derivatives $u_x$, $u_y$, $v_x$ and $v_y$.

The importation of the netCDF files in the ocean current model occurs within the NetCDF library [14]. All ocean current data will be stored in multidimensional array structures. For the interpolation a multi-dimensional interpolation scheme described in section III.A will be used. The usage of a different interpolation method for the two dimensional interpolation of the several depth layers and for the one dimensional interpolations of depth and time is possible.

Since all calculations in the simulator use SI units and a Cartesian coordinate system, an adaption (projection) to geodetic referenced coordinates from netCDF is necessary. This is the task of the proj4 library which supports a wide range of conversions between cartographic projections. For the two OGCM systems the following projections were used:

HYCOM:
```
"+proj=merc    +a=6371000.0    +b=6371000.0
+lat_ts=0.0 +lon_0=0.0 +x_0=0.0 +y_0=0 +k=1.0
+units=m +nadgrids=@null +no_defs"
```

CMEMS:
```
"+proj=eqc    +lat_ts=0    +lat_0=0    +lon_0=0
+x_0=0    +y_0=0    +ellps=WGS84    +datum=WGS84
+units=m +no_defs"
```

To analyze and to provide a graphical illustration of the particle tracks and sites, MATLAB will be used. The conversion of data of the XML files to MATLAB data types occurs by using the xml_io_tools library [15] and [16].

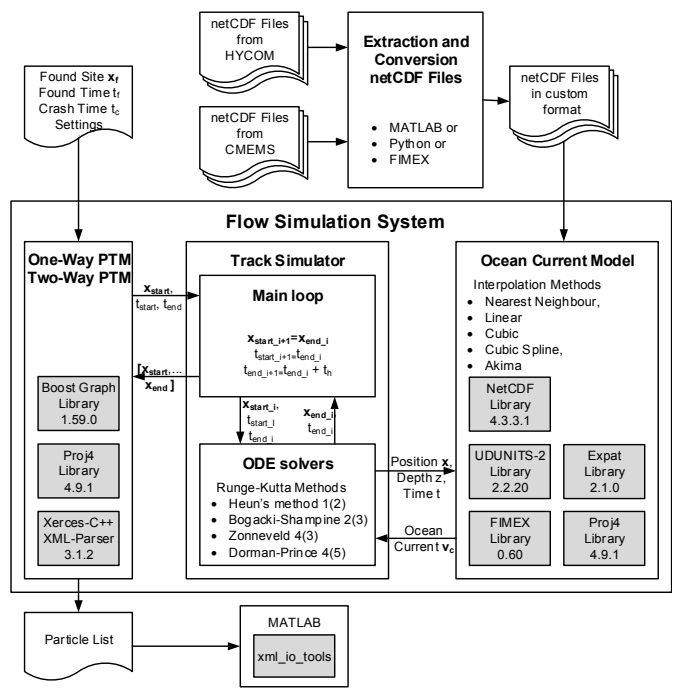

Fig. 3. Software concept of the flow simulation system

## V. RESULTS

This section presents the results of the flow simulation system. The first part of the section describes the tests to evaluate the simulation system using a mathematical model to generate time-variant ocean data. In the second part, the system will be used to analyze possible crash sites of MH370.

### A. Mathematical Ocean Model

The function to simulate a realistic time–varying ocean flow is based on a dimensionless function which describes a meandering jet in eastward direction, which is a simple mathematical model of the Gulf Stream [17] and [18]. The stream function is:

$$\phi(x,y) = 1 - \tanh\left( \frac{y - B(t)\cos(k(x-ct))}{\left(1 + k^2 B(t)^2 \sin^2(k(x-ct))\right)^{\frac{1}{2}}} \right) \quad (3)$$

It uses a dimensionless function of a time-dependent oscillation of the meander amplitude

$$B(t) = B_0 + \varepsilon \cos(\omega t + \theta) \quad (4)$$

and the parameter set $B_0 = 1.2$, $\varepsilon = 7.3$, $\omega = 0.4$, $\theta = \pi/2$, $k = 0.84$ and $c = 0.12$ to describe the velocity field:

$$u(x,y,t) = -\frac{\partial \phi}{\partial y} \quad v(x,y,t) = \frac{\partial \phi}{\partial x}. \quad (5)$$

To use this dimensionless function to simulate a realistic time-varying ocean flow the time scale corresponds to 3 days, and the space scales in x and y direction to 40 km.

The first test should analyse the stability and the accordance between the backward-in-time and forward-in-time simulation. For this, two crash sites are defined $x_1$ =[10 10] km; $x_2$ =[10 0] km with a distance of only 10 km between them. A Fifth-order Runge-Kutta method with step size control was used for all backward-in-time and forward in-time simulations. TABLE II shows the used parameters of the simulation. For these tests all algorithms were coded in MATLAB. This allows easy debugging and analysis as well as a comparison with the MATLAB ODE-solvers. After these successful tests the algorithms were rewritten in C++. The parameter time horizon $t_h$ includes the simulation time for the ODE-solver (see section IV). The ODE-solver ode113 from MATLAB was used to calculate a reference trajectory as well as the found site, which will be used for the backward-in-time calculation. This solver is an efficient multi step solver and is useful for problems with stringent error tolerances or computationally intensive ordinary differential equation functions [5]. Fig. 4 shows the calculated tracks for the backward-in-time simulation and for verifying the forward-in-time simulation.

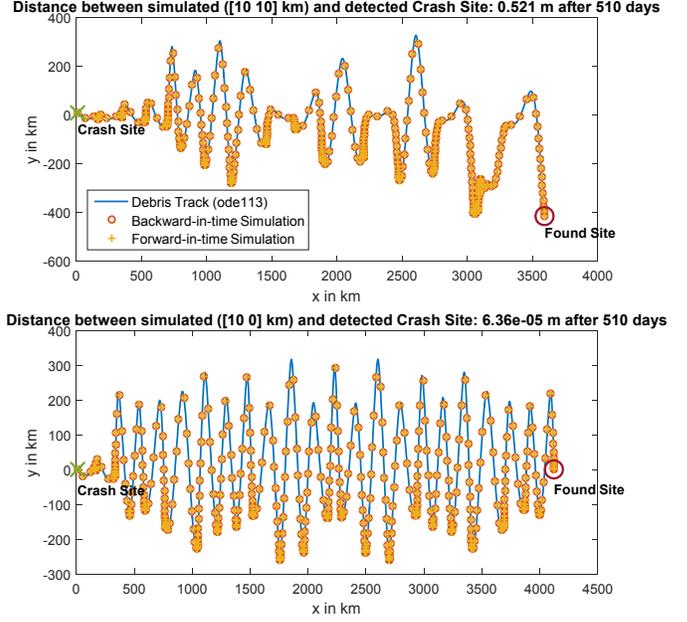

Fig. 4. Simulated tracks in backward- and forward-in-time simulation

The achieved accuracy between the simulated (start position of the ode113 solver) and the detected crash site (arrived position using the backward-in-time simulation starting from the found site) should not be overestimated. By using other start conditions the distance can be up to a few kilometers. It is also possible that the backward-in-time simulation cannot arrive near the crash position and stopped hundreds of kilometers away from it. Each marker in the diagrams shows the achieved position after 24 hours. An interesting fact is the completely different other trajectories and thus the found sites which are hundreds of kilometers away from each other. However the two defined crash sites are only 10 km away from each other. This behavior is comparable to the real ocean current field in the Indian Ocean. The usage of a Second-order Runge-Kutta method (see section II.B) often cannot solve the detection of the crash position with the defined test conditions.

### B. Comparison between the ODE Solvers

This section presents the results by using fixed step size ($h = h_{min}= 86.4$ s) and step size control for the several ODE solvers to solve the simulation task in the previous section. TABLE III shows the necessary current function calls. The step size control in the Fifth-order ODE leads to a performance-enhancement of over 90% compared to a second-order ODE, which is necessary to calculate hundreds of particle tracks in a Two-Way PTM in an acceptable time.

TABLE II. SIMULATION PARAMETERS

| Parameter | Value |
|---|---|
| Time horizon $t_h$ | 24 h |
| Tolerance $\varepsilon_{tol}$ | 0.0005 |
| $h_{min}$ | $t_h$ 0.001 = 86.4 s |
| $h_{max}$ | $t_h$ 0.1 = 2.4 h |

TABLE III. RESULTS OF THE DIFFERENT ODE SOLVERS

| Order | No. of Current Function Calls | |
|---|---|---|
| | Fixed Step Size | Step Size Control |
| 2 | 511020 | 481289 |
| 3 | 1532040 | 125142 |
| 4 | 2553060 | 97695 |
| 5 | 3063570 | 34416 |

## C. MH 370 Search

We use a modified version of the Two-Way PTM, which is described in section II.C. The used version works without a random walk process. The result of the random behavior of the several particle tracks to create possible crash sites are a randomly chosen start site on the red line in Fig. 5 (Marker1: 55.661850°E, 20.910733°S; Marker2: 55.684904°E, 20.933654°S) and a random chosen found time in a range from 2015-07-25 00:00:00 - 2015-07-28 00:00:00 UTC. The information about the found site and time is from [19] and [20]. We use four ocean current models to simulate the particle tracks over seventeen month in the Indian Ocean. These are, GLBa0.08 (HYCOM GLBa008) and GLBu0.08 (HYCOM GLBu008) of HYCOM [3] and the 2 hourly (CMEMS 2HOURLY) and 24 hourly (CMEMS 24HOURLY) global-analysis-forecast-phys-001-002 model of the Copernicus Marine environment monitoring service (CMEMS) [4]. For all models only the surface current respectively the first depth layer was used. This means for HYCOM z=0 m and for CMEMS z=0.4940254 m. The area of interest is 54°E-117°E 42°S-10°N for both systems in a time range from March 2014 to August 2015.

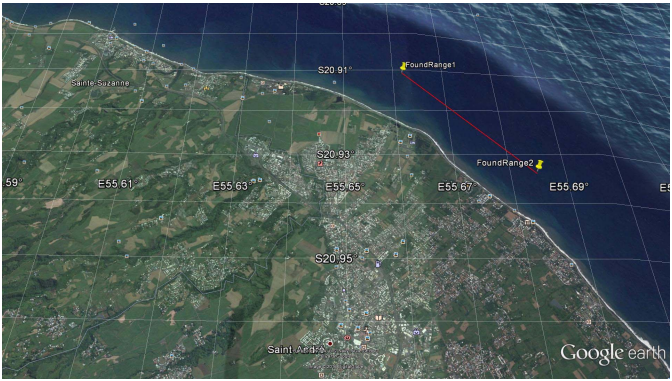

Fig. 5. Definition of the start region near the beach in St Andre, on the north-eastern coast of La Réunion

A Fifth-order Runge-Kutta method with step size control was used for all backward-in-time and forward-in-time simulations of the PTMs. The simulation parameters are set according to the preview test in TABLE II. For all interpolations in the ocean current model Cubic Spline interpolation was used. The program runs on a Window 7 64 bit operation system on a Dell Precision M4800 Laptop with a Quad Core Intel i7-4900MQ and 32 GB Memory. This memory size was necessary to load netCDF files of the CMEMS 2HOURLY model up to a size of 23 GB.

To create possible crash site candidates a backward-in-time PTM with 1000 particles for each model was run which is the first step of the two-way PTM. Fig. 6 shows the possible crash sites using all four models. The several steps of the two-way PTM will be demonstrated on the two CMEMS models following. Fig. 7 shows the possible crash sites using CMEMS 24HOURLY and the associated histogram in Fig. 8. The defined sector elements have a size of 2.5°Lon and 2.5°Lat.

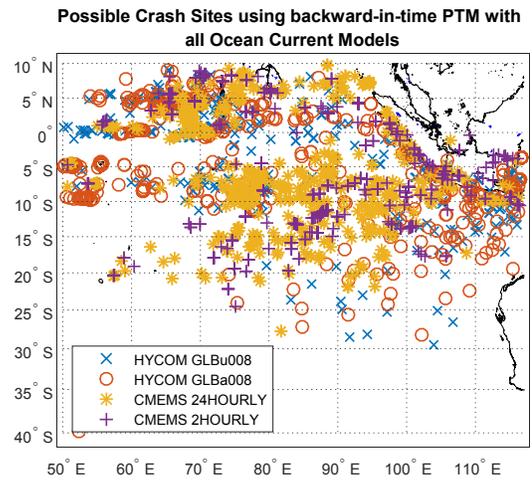

Fig. 6. Possible crash sites using all four ocean current models

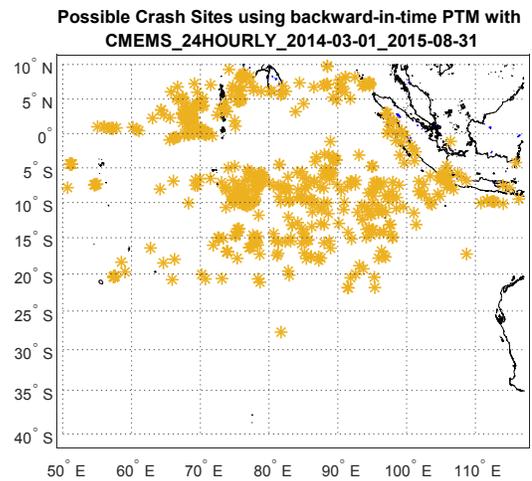

Fig. 7. Possible crash sites using backward-in-time PTM with CMEMS 24HOURLY (First Step of the Two-Way PTM)

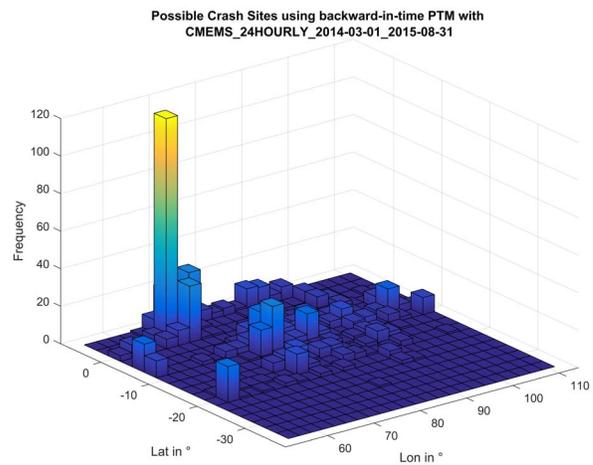

Fig. 8. Histogram of possible crash sites. Select possible crash site candidates on the basis of numbers of particles in the several sectors

Three possible crash sites were chosen by means of the histogram in Fig. 8. The sites were calculated by using all particle positions in the chosen sectors. The calculated average positions are shown in TABLE IV.

TABLE IV. RESULTS OF THE TWO-WAY PTM USING CMEMS 24HOURLY

| Name | Possible Crash Sites Average Position of all Particles in Sector | Found Sites Average Position of all Particles |
|---|---|---|
| Source 1 | 68.0°E, 2.6°N | 68.6°E, 21.7°S |
| Source 2 | 77.2°E, 9.9°S | 55.7°E, 20.4°S |
| Source 3 | 105.5°E, 6.7°S | 78.9°E, 24.6°S |
| Hot-Spot | 88°E, 38°S | 95.0°E, 30.9°S |

As an additional possible crash site the hot-spot which was defined in [21] was used. 250 particles for each of the four crash sites were created by a random variation within a radius of 3 km and a random variation of the crash time of 2 hours from an assumed crash time at 2014-03-08 00:19:00 UTC. Fig. 9 shows the result of the possible found sites for the several crash site candidates using a forward-in-time PTM which is the second step of the two-way PTM. The average position of all particles is marked with a black x in a colored circle. In this simulation Source 2 was favored as a possible crash site, because the average position of all particles is the closest to La Réunion. This result has to be verified because only particles which are inside of the area of interest were used for the analyses. Several particles flew more westwards and went outside of the area of interest. In such cases the interpolation algorithms work with a rough approximation of the ocean current field. Fig. 10 shows the histogram of the particles starting from Source 2.

The mechanism of the two-way PTM as described above will now be repeated for the CMEMS 2HOURLY model. The results also show a favored crash site of Source 2 (see TABLE V) which is near the first Source 2 position using CMEMS 24HOURLY. A verification of the result is also necessary, because several particles went outside of the area of interest.

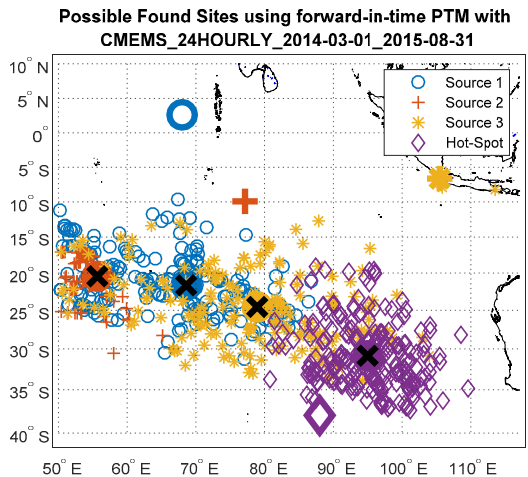

Fig. 9. Possible found sites using forward-in-time PTM with CMEMS 24HOURLY (Second Step of the Two-Way PTM)

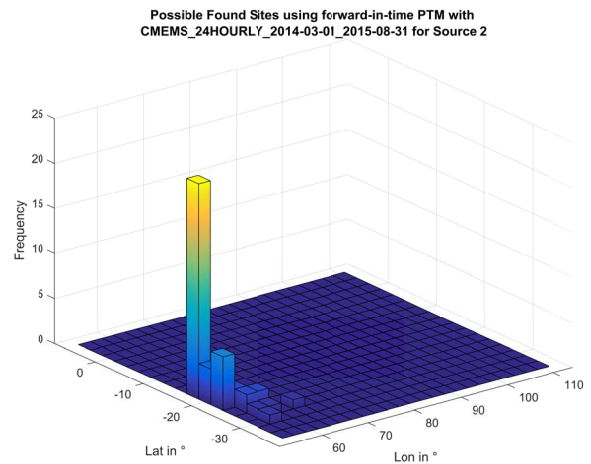

Fig. 10. Histogram of possible found sites for Source 2

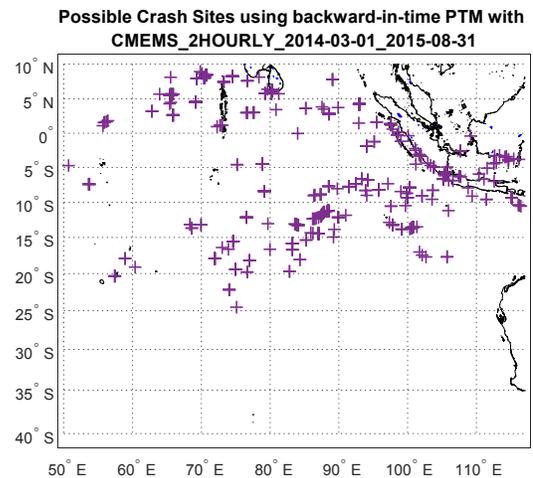

Fig. 11. Possible crash sites using backward-in-time PTM with CMEMS 2HOURLY (First Step of the Two-Way PTM)

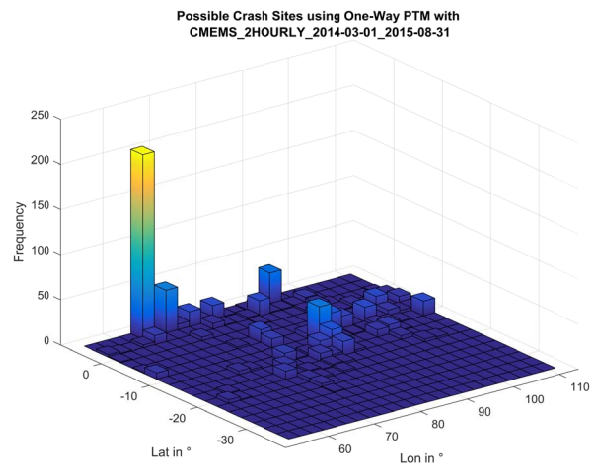

Fig. 12. Histogram of possible crash sites. Select possible crash site candidates on the basis of numbers of particles in the several sectors

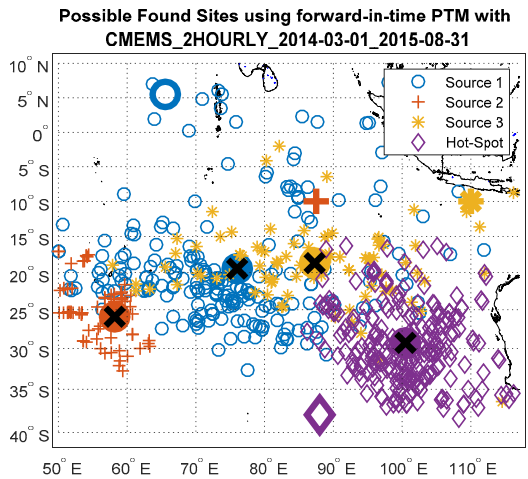

Fig. 13. Possible found sites using forward-in-time PTM with CMEMS 2HOURLY (Second Step of the Two-Way PTM)

TABLE V. RESULTS OF THE TWO-WAY PTM USING CMEMS 2HOURLY

| Name | Possible Crash Sites Average Position of all Particles in Sector | Found Sites Average Position of all Particles |
|---|---|---|
| Source 1 | 65.6°E, 5.5°N | 76.0°E, 19.5°S |
| Source 2 | 88.3°E, 10.8°S | 58.2°E, 25.9°S |
| Source 3 | 114.6°E, 9.6°S | 87.3°E, 18.7°S |
| Hot-Spot | 88°E, 38°S | 100.5°E, 29.3°S |

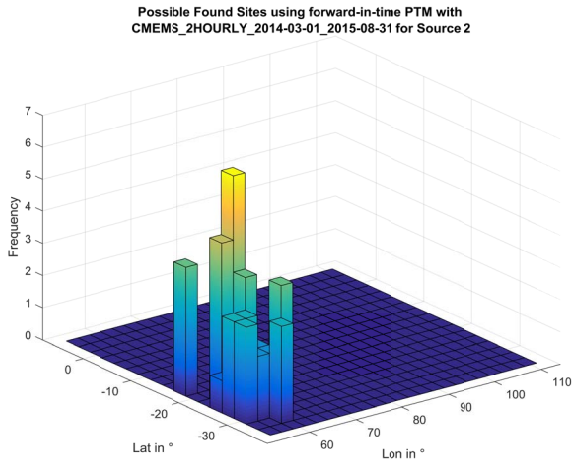

Fig. 14. Histogram of possible found sites for Source 2

## VI. CONCLUSIONS

This work presented a system to analyze possible crash positions. A two-way Particle Tracking Model was used to simulate the flow of the found debris of MH370 using a simple flow simulation based on a Lagrangian approach. For this purpose, ocean general circulation models provided the necessary ocean current data. A software framework for ocean data extraction and preparation was introduced. The simulation system was evaluated using a mathematical ocean model. Finally, the system was used to analyze possible crash sites of MH370.

In conclusion, the used ocean current models have a large uncertainty over such a long time period. The area of interest, the Indian Ocean, has a complex flow structure characterized by eddies and tidal ocean currents in shallow water areas. Furthermore, the HYCOM and CMEMS models show a different ocean current behavior in a few areas and time ranges. All these facts should be taken into consideration when interpreting the results. The additional debris which were found recently could help to confirm or to reject the presented results. Also, the inclusion of additional information from drifting buoys [22] or the region of provenance of the attached barnacles using statistics methods as in [21] could help to solve this puzzle.